\documentclass[aps,prd,preprint,amsmath,amssymb]{revtex4}
\usepackage{graphicx}
\usepackage{dcolumn}
\usepackage{bm}

\newcommand{\be}{\begin{equation*}}
\newcommand{\ee}{\end{equation*}}
\newcommand{\bea}{\begin{eqnarray}}
\newcommand{\eea}{\end{eqnarray}}
\newcommand{\bean}{\begin{eqnarray*}}
\newcommand{\eean}{\end{eqnarray*}}
\newcommand{\ga}{{a^0_0(980)}}
\newcommand{\gf}{{f_0(980)}}
\newcommand{\gaf}{{$a^0_0(980)$-$f_0(980)$ mixing\ }}

\begin{document}

\title{Study \gaf from $\ga\to\gf$ transition}

\author{
Jia-Jun Wu$^{1,2}$ and B.~S.~Zou$^{1,2}$\\
$^1$ Institute of High Energy Physics, CAS, P.O.Box 918(4), Beijing 100049, China\\
$^2$ Theoretical Physics Center for Science Facilities, CAS, Beijing
100049, China}

\date{August 19, 2008}

\begin{abstract}
Various processes have been proposed previously to study \gaf
through $\gf\to\ga$ transition. Here we investigate in detail the
difference between $\ga\to\gf$ and $\gf\to\ga$ transitions. It is
found that the $\ga\to\gf$ transition can provide additional
constrains to the parameters of $\ga$ and $\gf$ mesons. Proposal is
made to study \gaf from $\chi_{c1}\to\pi^{0}\ga\to\pi^{0}\gf$
reaction at the upgraded Beijing electron positron collider with the
BESIII detector.
\end{abstract}


\maketitle

\section{Introduction}
\label{s1}

More than thirty years after their discovery, today the nature of
light scalar mesons $\gf$ and $\ga$ is still in controversy. They
have been described as quark-antiquark, four quarks,
$K\bar{K}$molecule, quark-antiquark-gluon hybrid, and so on. Now the
study of their nature has become a central problem in the light
hadron spectroscopy.

In the late 1970s, the mixing between the $\ga$ and $\gf$ resonances
was first suggested theoretically in Ref.\cite{first}. Its mixing
intensity is expected to shed important light on the nature of these
two resonances, and has hence been studied extensively on its
different aspects and possible manifestations in various reactions
\cite{n1,n2,n3,n5,im,uim,n4,y2,y4,y5,y10,y11,eb,we,hanhart}. But
unfortunately no firm experimental determination on this quantity is
available yet. Obviously, more solid and precise measurements on
this quantity are needed, such as by polarized target experiment on
the reaction $\pi^-p\to\eta\pi^0 n$ \cite{n3}, $J/\psi$ decays
\cite{im,we}, and $dd\to\alpha\eta\pi^0$ reactions from WASA at COSY
\cite{y10}. In Ref.~\cite{we}, we pointed out that the \gaf
intensity can be precisely measured through
$J/\psi\to\phi\gf\to\phi\ga\to\phi\eta\pi^0$ reaction at the
upgraded Beijing electron positron collider with the BESIII
detector.

In all these previous proposals, the $\gf$ is produced first, then
transits to the $\ga$ by the \gaf, {\sl i.e.}, $\gf\to\ga$
transition. In this article, we investigate in detail the difference
between $\ga\to\gf$ and $\gf\to\ga$ transitions. We define two kinds
of mixing intensities $\xi_{fa}$ and $\xi_{af}$ for the $\gf\to\ga$
and $\ga\to\gf$ transitions, respectively. We find there are some
differences between them. We find that $\xi_{af}$ has more
dependence on the parameters of $\gf$, especially the
$g_{fKK}/g_{f\pi\pi}$, while the $\xi_{fa}$ has more dependence on
the parameters of $\ga$, especially $g_{aKK}/g_{a\pi\eta}$. For this
reason, using the reaction of
$J/\psi\to\phi\gf\to\phi\ga\to\phi\eta\pi^0$ to study only the
$\gf\to\ga$ mixing is not perfect enough. For better determination
of all relevant parameters for the $\gf$ and $\ga$ mesons, it would
be useful to find some reaction to study $\ga\to\gf$ mixing in
addition. Recently, CLEO Collaboration reported an experimental
study of the $\chi_{c1} \to \pi^+ \pi^-\eta$ reaction~\cite{cleo}.
The $a^{\pm}_0(980)$ resonances are clearly showing up and dominant.
>From isospin symmetry, the $\chi_{c1}\to\pi^0\ga$ should be produced
with the same rate as $\chi_{c1}\to\pi^\pm a^{\mp}_0(980)$. This may
provide a nice place for studying the \gaf from $\ga\to\gf$
transition by $\chi_{c1}\to\pi^{0}\ga\to\pi^{0}\gf\to \pi^{0}\pi\pi$
reaction. From our estimation, more than 300 events can be
reconstructed by the BESIII detector in the narrow peak with a width
of about 8 MeV around the mass of 990 MeV in the $\pi\pi$ invariant
mass spectrum.

In the next section, we give a brief review of the theory for the
\gaf term. Then in the Sect.\ref{s3} we define two mixing
intensities and tell the differences of mixing intensity. In
Sect.\ref{s5} we estimate the rate for the reaction of $\chi_{c1}\to
\pi^{0}\pi\pi$. Finally we give a summary in Sect.\ref{s6}.

\section{The $a^0_0(980)$-$f_0(980)$ Mixing Amplitude}
\label{s2}

The basic theory for the \gaf was already pointed out by Achasov
and collaborators \cite{first}. For the nearly degenerate $\ga$
(isospin 1) and $\gf$ (isospin 0), both can decay into $K\bar K$.
Due to isospin breaking effect, the charged and neutral kaon
thresholds are different by about 8 MeV. Between the charged and
neutral kaon thresholds the leading term to the \gaf amplitude is
dominated by the unitary cuts of the intermediate two-kaon system
and proportional to the difference of phase spaces for the charged
and neutral kaon systems.

Considering the \gaf, the propagator of $\ga/\gf$ can be expressed
as \cite{n1} :
\begin{eqnarray}
G=\frac{1}{D_{f}D_{a}-|D_{af}|^{2}}\begin{pmatrix}D_{a}&D_{af}\\D_{af}&D_{f}\end{pmatrix},\label{1}
\end{eqnarray}
where $D_{a}$ and $D_{f}$ are the denominators for the usual
propagators of $\ga$ and $\gf$, respectively :
\begin{eqnarray}
D_{a}&=&m_{a}^{2}-s- i\sqrt{s}[\Gamma^a_{\eta\pi}(s) +
\Gamma^a_{K\bar K}(s)],\\\label{2} D_{f}&=&m_{f}^{2}-s-
i\sqrt{s}[\Gamma^f_{\pi\pi}(s)+\Gamma^f_{K\bar K}(s)],\\\label{3}
\Gamma^a_{bc}(s)&=&\frac{g^{2}_{abc}}{16\pi\sqrt{s}}\rho_{bc}(s),\\\label{4}
\rho_{bc}(s)&=&\sqrt{[1-(m_{b}-m_{c})^{2}/s][1-(m_{b}+m_{c})^{2}/s]}\label{5}.
\end{eqnarray}
The  $D_{af}$ is the mixing term. From \cite{first,n3}, the mixing
due to $K\bar K$ loops gives
\begin{eqnarray}
D_{af,K\overline{K}}&=&\frac{g_{\ga K^{+}K^{-}}g_{\gf K^{+}K^{-}}}{16\pi}
\Big\{i[\rho_{K^{+}K^{-}}(s)-\rho_{K^{0}\bar{K}^0}(s)]\nonumber\\
&&-\mathcal{O}(\rho^{2}_{K^{+}K^-}(s)-\rho^{2}_{K^{0}\bar{K}^0}(s))\Big\}.\label{6}
\end{eqnarray}

Since the mixing is mainly coming from the $K\bar K$ loops, we have
$D_{af}\approx D_{af,K\bar K}$, and this is the amplitude of \gaf.
>From Eq.(\ref{6}), the $D_{af}$ becomes large only when the
$\sqrt{s}$ is between the $2M_{K^{+}}$ and $2M_{K^{0}}$, so it is a
narrow peak with the width of about $8MeV$.

\section{Two types of reaction and mixing intensity }
\label{s3}

\subsection{Two types of reaction of \gaf }

There are two types of reaction which can be used to study \gaf: $X
\to Y \gf \to Y \ga \to Y \pi^{0}\eta$ and $X \to Y \ga \to Y \gf
\to Y \pi\pi$.

\begin{figure}[htbp] \vspace{-0.cm}
\begin{center}
\includegraphics[width=0.55\columnwidth]{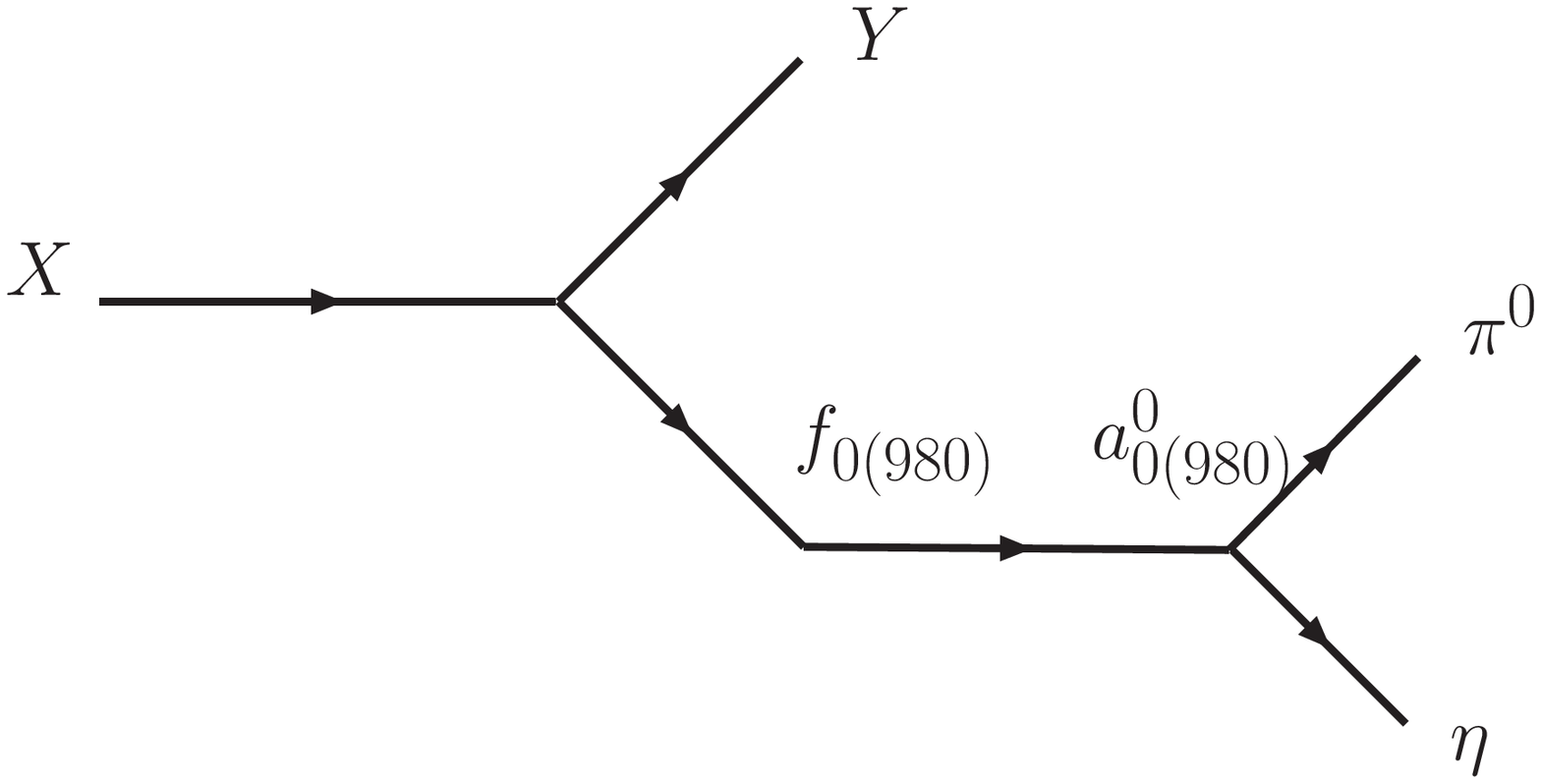}
\caption{The Feynman diagram of $X \to Y \gf \to Y \ga \to Y
\pi^{0}\eta$.} \label{fg1}
\end{center}
\end{figure}

\begin{figure}[htbp] \vspace{-0.cm}
\begin{center}
\includegraphics[width=0.55\columnwidth]{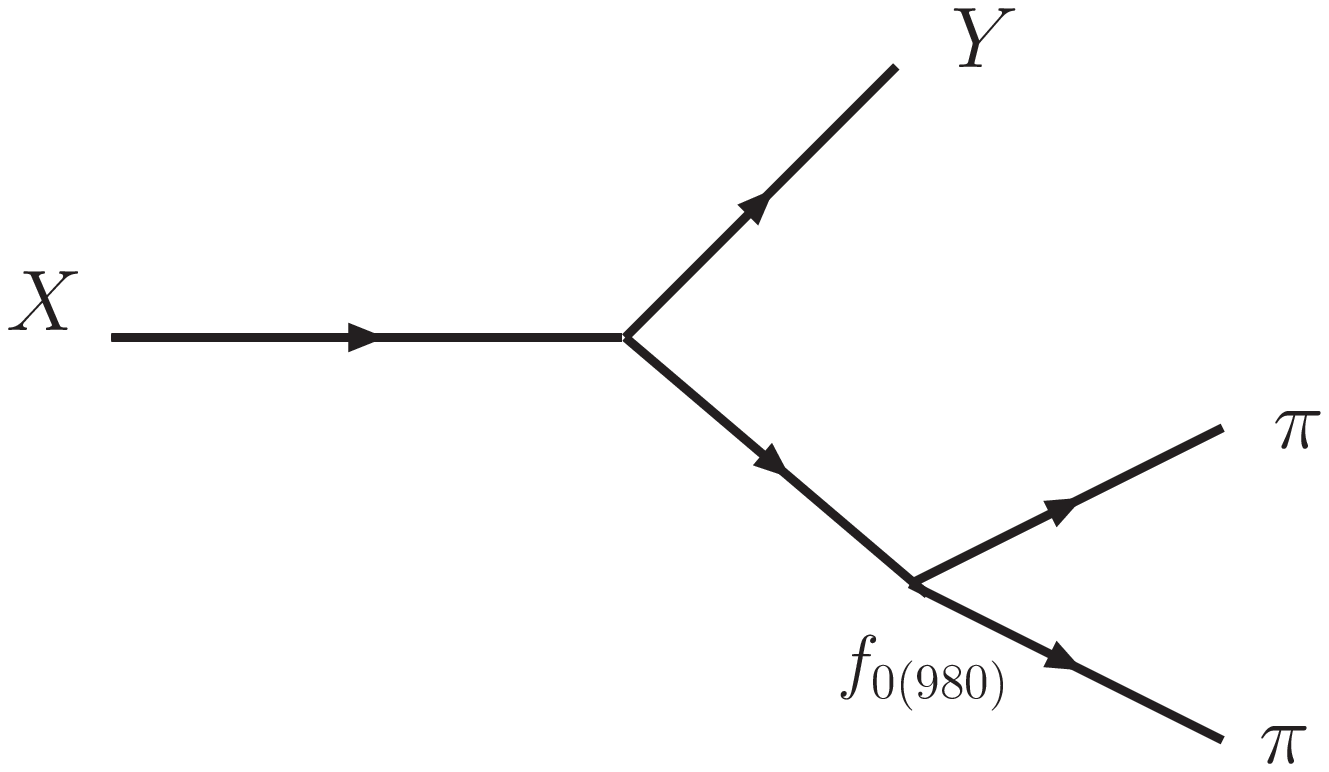}
\caption{The Feynman diagram of $X \to Y \gf \to Y  \pi\pi$.}
\label{fg2}
\end{center}
\end{figure}

For the reaction $X \to Y \gf \to Y \ga \to Y \pi^{0}\eta$ as shown
by the Feynman diagram in Fig.\ref{fg1}, the influence of various X
and Y on the \gaf can be removed by its comparison to the
corresponding reaction $X \to Y \gf \to Y \pi\pi$ as shown in
Fig.\ref{fg2}. We define the mixing intensity $\xi_{fa}$ for the
$\gf\to\ga$ transition as the following
\begin{equation}
\xi_{fa}(s)=\frac{d\Gamma_{X \to Y \gf \to Y \ga \to Y
\pi^{0}\eta(s)}}{d\Gamma_{X \to Y \gf \to Y \pi\pi(s)}}\label{7},
\end{equation}
where $s$ is the invariant mass squared of two mesons in final
state. With Eqs.(1-6), one can get the $\xi_{fa}(s)$ as:
\begin{eqnarray}
\xi_{fa}(s)&=&\frac{|D_{af}|^{2}\Gamma^{a}_{\pi\eta}}{|D_{a}|^{2}\Gamma^{f}_{\pi\pi}}\\
&=&\left|\frac{g_{a^{0}_0K^{+}K^{-}}g_{f_{0}K^{+}K^{-}}}
{g_{a^{0}_{0}\pi^{0}\eta}g_{f_{0}\pi^{0}\pi^{0}}}\right|^{2}
\frac{\left|\rho_{K^{+}K^{-}}(s)-\rho_{K^{0}\bar{K}^0}(s)\right|^{2}}
{3\rho_{\pi\pi}(s)\rho_{\pi\eta}(s)}\times\nonumber\\
&&\frac{1}{\left|\frac{m^{2}_{a}-s}{\Gamma^{a}_{\pi\eta}\sqrt{s}}
-i\left[\left|\frac{g_{a^{0}_{0}K^{+}K^{-}}}{g_{a^{0}_{0}\pi^{0}\eta}}\right|^{2}
\left(\frac{\rho_{K^{+}K^{-}(s)}}{\rho_{\pi\eta(s)}}+\frac{\rho_{K^{0}\bar{K}^{0}}}
{\rho_{\pi\eta}}\right)+1\right]\right|^{2}}.\label{9}
\end{eqnarray}

Similarly, for the reaction $X \to Y \ga \to Y \gf \to Y \pi\pi$, we
define the mixing intensity $\xi_{af}$ for the $\ga\to\gf$
transition and get its formula as the following
\begin{eqnarray}
\xi_{af}(s)&=&\frac{d\Gamma_{X \to Y \ga \to Y \gf \to Y
\pi\pi(s)}}{d\Gamma_{X \to Y \ga \to Y \pi^0\eta(s)}},\\
&=&\left|\frac{g_{a^{0}_{0}K^{+}K^{-}}g_{f_{0}K^{+}K^{-}}}
{g_{a^{0}_{0}\pi^{0}\eta}g_{f_{0}\pi^{0}\pi^{0}}}\right|^{2}
\frac{\left|\rho_{K^{+}K^{-}}(s)-\rho_{K^{0}\bar{K}^0}(s)\right|^{2}}
{3\rho_{\pi\pi}(s)\rho_{\pi\eta}(s)}\times\nonumber\\
&&\frac{1}{\left|\frac{m^{2}_{f}-s}{\Gamma^{f}_{\pi\pi}\sqrt{s}}
-i\left[\left|\frac{g_{f_{0}K^{+}K^{-}}}{g_{f_{0}\pi^{0}\pi^{0}}}\right|^{2}
\left(\frac{\rho_{K^{+}K^{-}(s)}}{3\rho_{\pi\pi(s)}}
+\frac{\rho_{K^{0}\bar{K}^{0}(s)}}{3\rho_{\pi\pi(s)}}\right)+1\right]\right|^{2}}.\label{10}
\end{eqnarray}

We can redefine that:
\begin{eqnarray}
&R_{f}=\left|g_{f_{0(980)}K^{+}K^{-}}/ g_{f_{0(980)}\pi^{0}\pi^{0}}\right|^{2}&,
\quad R_{a}=\left|g_{a^{0}_{0(980)}K^{+}K^{-}}/ g_{a^{0}_{0(980)}\pi^{0}\eta}\right|^{2},\label{pa1}\\
&A^f_{(s)}=\rho_{K^{+}K^{-}(s)}/ 3\rho_{\pi\pi(s)}&,
\quad A^a_{(s)}=\rho_{K^{+}K^{-}(s)}/ \rho_{\pi\eta(s)},\\
&B^f_{(s)}=\rho_{K^{0}\bar{K}^{0}(s)}/ 3\rho_{\pi\pi(s)}&,
\quad B^a_{(s)}=\rho_{K^{0}\bar{K}^{0}(s)}/ \rho_{\pi\eta(s)},\\
&C^f_{(s)}=(m^{2}_{f}-s)/ \Gamma^{f}_{\pi\pi}\sqrt{s}&, \quad
C^a_{(s)}=(m^{2}_{a}-s)/
\Gamma^{a}_{\pi\eta}\sqrt{s},\\
&H_{(s)}=\left|\rho_{K^{+}K^{-}}(s)-\rho_{K^{0}\bar{K}^0}(s)\right|^{2}&/
\left(3\rho_{\pi\pi}(s)\rho_{\pi\eta}(s)\right).\label{pa2}
\end{eqnarray}

The $D_{af}$ becomes large when the $\sqrt{s}$ is between the
$2M_{K^{+}}$ and $2M_{K^{0}}$. In this mass range, $A^f_{(s)}$ and
$A^a_{(s)}$ are real, meanwhile $B^f_{(s)}$ and $B^a_{(s)}$ are
imaginary; then $\xi_{af}$ and $\xi_{fa}$ become:
\begin{eqnarray}
\xi_{af}(s)&=&\frac{R_{a}R_{f} \times H_{(s)}}{\left(C^f_{(s)}+R_{f}\times |B^f_{(s)}|\right)^{2}
+\left(1+R_{f} \times A^f_{(s)}\right)^{2}}\label{y1} ,\\
\xi_{fa}(s)&=&\frac{R_{a}R_{f} \times
H_{(s)}}{\left(C^a_{(s)}+R_{a}\times
|B^a_{(s)}|\right)^{2}+\left(1+R_{a}\times A^a_{(s)}\right)^{2}}
.\label{y2}
\end{eqnarray}

\subsection{Predictions of $\xi_{af}$ and $\xi_{fa}$ from various models and experiment information}

\begin{figure}[htbp] \vspace{-0.cm}
\begin{center}
\includegraphics[width=0.48\columnwidth]{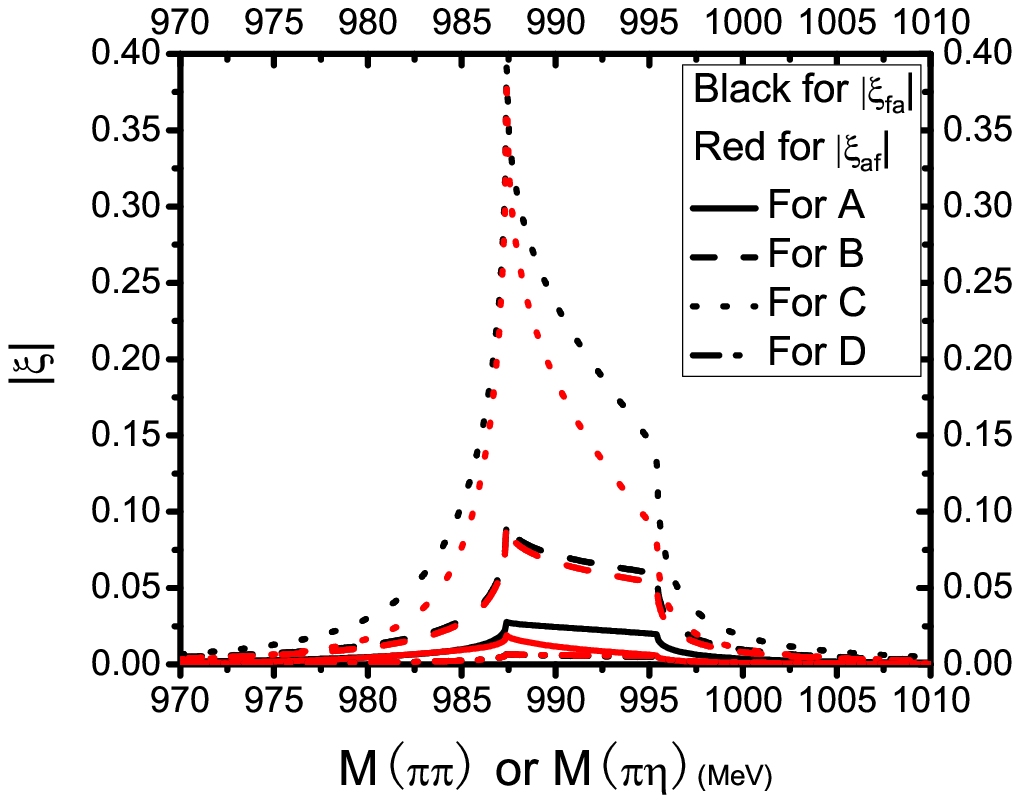}
\includegraphics[width=0.48\columnwidth]{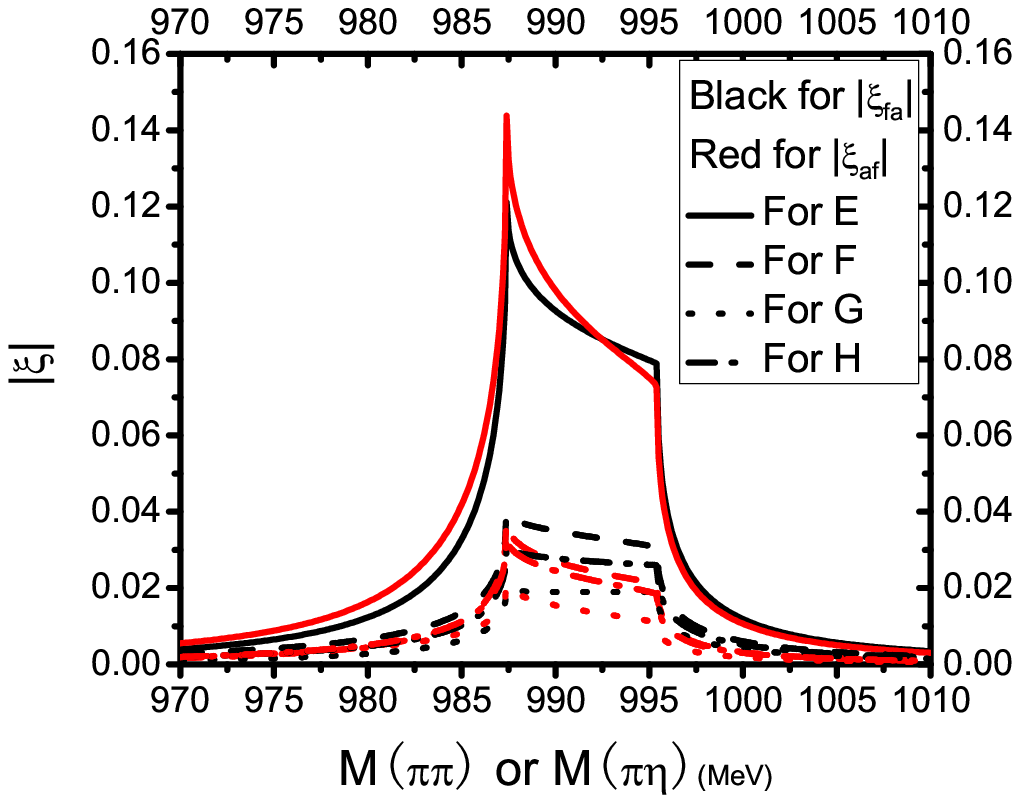}
\caption{Predictions for the \gaf intensity $\xi_{af}$ and
$\xi_{fa}$ vs two-meson invariant mass from various models A-D
(left) and various experimental measured parameters E-H (right).}
\label{f1}
\end{center}
\end{figure}

>From equations given above, one can see that the mixing intensity
$|\xi|$ depends on $g_{\ga K^{+}K^{-}}$, $g_{\gf K^{+}K^{-}}$,
$g_{\ga\pi^{0}\eta}$, $g_{\gf\pi^{0}\pi^{0}}$, $m_{f}$ and $m_{a}$.
Various models for the structures of $\ga$ and $\gf$ give different
predictions for these coupling constants and mass \cite{eb,m2,m3,m6}
as listed in Table \ref{tab-1} by No.A-D. There have also been some
experimental measurements on these coupling and mass constants
\cite{ge1,ge2,ge4,ge5,m5,zou1,zou2,bes1} as listed by No.E-H. The
corresponding predictions for the $\xi_{af}$ and $\xi_{fa}$ from
these various theoretical and experimental values of the coupling
constants are calculated. In the calculation, the masses for $K^+$,
$K^0$, $\pi^0$ and $\eta$ are taken from PDG2008 \cite{pdg06} as
$m_{K^{+}}=493.7$ MeV, $m_{K^{0}}=497.7$ MeV, $m_{\pi}=135.0$ MeV
and $m_{\eta}=547.5$ MeV, respectively. We give the value of
$\xi_{af}$ and $\xi_{fa}$ at the $\sqrt{s}=991.4$ MeV in Table
\ref{tab-1} and the dependence of \gaf intensities $\xi_{af}$ and
$\xi_{fa}$ vs two-meson invariant mass in Fig.(\ref{f1}). There is
obviously some difference between these two mixing intensities.

\begin{table}[ht]
\begin{tabular}{|c|c|c|c|c|c|c|c|c|c|}
\hline No. & model or experiment & $m_{a}$ & $g_{a_{0}\pi\eta}$ &
$g_{a_{0}K^{+}K^{-}}$ & $m_f$ & $g_{f_{0}\pi^{0}\pi^{0}}$ & $g_{f_{0}K^{+}K^{-}}$ & $|\xi_{fa}|$ & $|\xi_{af}|$\\
\hline
 A & $q\bar{q}$ model \cite{eb} & 983 & 2.03 & 1.27 & 975 & 0.64 & 1.80 & 0.023 & 0.010\\
\hline
 B & $q^{2}\bar{q}^{2}$ model \cite{eb} & 983 & 4.57 & 5.37& 975 & 1.90 & 5.37 & 0.068 & 0.062\\
\hline
 C & $K\bar{K}$ model \cite{m2,m6,kk} & 980 & 1.74 & 2.74& 980 & 0.65 & 2.74 & 0.21 & 0.15\\
\hline
 D & $q\bar{q}g$ model \cite{m3} & 980 & 2.52& 1.97& 975 & 1.54 & 1.70 & 0.005 & 0.006\\
\hline
 E & SND \cite{ge1,ge2} & 995 & 3.11 & 4.20& 969.8 & 1.84 & 5.57 & 0.088 & 0.089\\
\hline
 F & KLOE \cite{ge4,ge5} & 984.8 & 3.02 & 2.24& 973 & 2.09 & 5.92 & 0.034 & 0.025\\
\hline
 G & BNL \cite{m5} & 1001 & 2.47 & 1.67& 953.5\cite{zou2} & 1.36\cite{zou2} & 3.26\cite{zou2} & 0.019 & 0.014\\
\hline
 H & CB \cite{zou1} & 999 & 3.33 & 2.54 & 965\cite{bes1} & 1.66\cite{bes1} & 4.18 \cite{bes1} & 0.027 & 0.023\\
\hline
\end{tabular}
\caption{$m_{\ga}$(MeV), $m_{\gf}$(MeV) and coupling constants
$g_{a_{0}\pi\eta}$(GeV), $g_{a_{0}K^{+}K^{-}}$(GeV),
$g_{f_{0}K^{+}K^{-}}$(GeV) and $g_{f_{0}\pi^{0}\pi^{0}}$(GeV) from
various models (A-D) and experimental measurements (E-H), and
calculated values of $|\xi_{af}|$ and $|\xi_{fa}|$ at the
$\sqrt{s}=991.4$ MeV by Eqs.(17,18).} \label{tab-1}
\end{table}

\subsection{Discussion on the difference of two mixing intensities $\xi_{af}$ and $\xi_{fa}$}

We know from Eqs.(17,18), if the $C^{i}_{(s)}$, $R_i\times
B^i_{(s)}$ and $R_{i} \times A^i_{(s)}$ (i=a or f) are much smaller
than 1, we will have $\xi_{af}(s)\simeq\xi_{fa}(s)\simeq R_{a}R_{f}
\times H_{(s)}$. Then from these mixing intensities we only get the
$R_{a}R_{f}$ no matter which type of reactions one makes the
measurement.

However, are these quantities really much smaller than 1 ? For
$\sqrt{s}$ between the $2M_{K^{+}}$ and $2M_{K^{0}}$,
$\sqrt{s}\approx m_i$, we have
\begin{equation}
C^{i}_{(s)}=(m^{2}_{i}-s)/
(\Gamma^{i}_{\pi\pi(\eta)}\sqrt{s})=(m_{i}-\sqrt{s})(m_{i}+\sqrt{s})/(\Gamma^{i}_{\pi\pi(\eta)}\sqrt{s})\approx
2(m_{i}-\sqrt{s})/\Gamma^{i}_{\pi\pi(\eta)}.\label{19}
\end{equation}
Here $\Gamma^{i}_{\pi\pi(\eta)}$ is about 100 MeV from many
experiments, $m_{i}\simeq 980$ MeV,  then one gets
$C^{i}_{(4M_{K^+}^{2})}=-0.14$ and $C^{i}_{(4M_{K^0}^{2})}=-0.30$,
which are not so small.

For $R_i\times B^i_{(s)}$ and $R_i \times A^i_{(s)}$ with
$A^{i}_{(4M_{K^+}^{2})}=|B^i_{(4M_{K^0}^{2})}|=0$,
$A^{a}_{(4M_{K^0}^{2})}\simeq|B^{a}_{(4M_{K^+}^{2})}|=0.201$ and
$A^{f}_{(4M_{K^0}^{2})}\simeq|B^{f}_{(4M_{K^+}^{2})}|=0.044$, since
from Table \ref{tab-1} the smallest $R_a$ and $R_f$ are 0.5 and 5.7
respectively among various experimental determinations, the
$R_i\times |B^i_{(s)}|$ and $R_i \times A^i_{(s)}$ are also larger
than 0.1 and are not small enough to be neglected.

Then from Eqs.(17,18), one can see that besides the common
numerator, the $\xi_{af}$ has additional dependence on the
parameters of $\gf$ while the $\xi_{fa}$ has additional dependence
on the parameters of $\ga$.

>From above analysis, we understand why $\xi_{af}$ is different from
$\xi_{fa}$ as shown in Table \ref{tab-1} and Fig.\ref{f1}. Both
mixing intensities depend on six parameters: $m_f$, $m_a$, $g_{\ga
K^{+}K^{-}}$, $g_{\gf K^{+}K^{-}}$, $g_{\ga\pi^{0}\eta}$,
$g_{\gf\pi^{0}\pi^{0}}$, which are all important for understanding
the nature of the $\ga$ and $\gf$ mesons, but are not well
determined yet.  Therefore to measure $\xi_{af}$ in addition to
$\xi_{fa}$ will be very useful for pinning down these parameters.

To further demonstrate the importance of measuring the $\xi_{af}$ in
addition to the $\xi_{fa}$, a typical example is given as the
following.

\begin{table}[ht]
\begin{tabular}{|c|c|c|c|c|c|c|c|c|c|}
\hline No.  & $m_{a}$ (MeV) & $g_{a_{0}\pi\eta}$ (GeV) &
$g_{a_{0}K^{+}K^{-}}$ (GeV) & $m_f$ (MeV) & $g_{f_{0}\pi^{0}\pi^{0}}$ (GeV) & $g_{f_{0}K^{+}K^{-}}$ (GeV)\\
\hline
 1  & 980 & 3.2 & 4.2& 980 & 1.5 & 4.0 \\
\hline
 2  & 980 & 3.2& 3.0& 980 & 1.5 & 5.12 \\
\hline
\end{tabular}
\caption{Two typical parameter sets for $\ga$ and $\gf$.}
\label{tab-2}
\end{table}

\begin{figure}[htbp] \vspace{-0cm}
\begin{center}
\includegraphics[width=0.48\columnwidth]{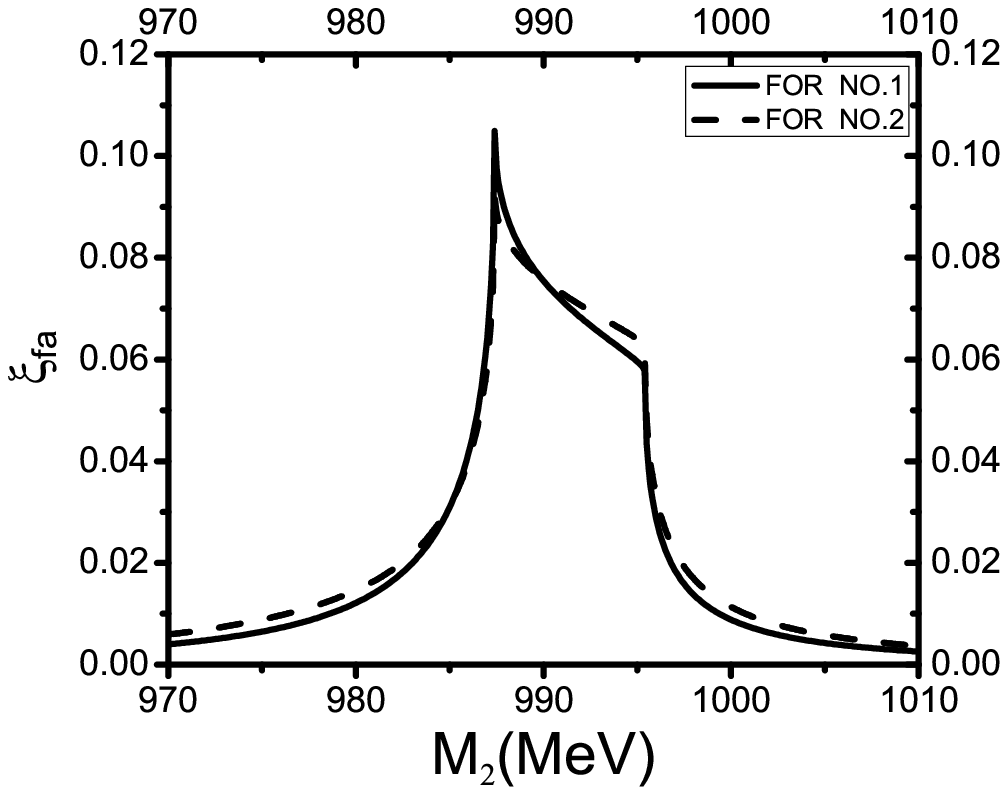}
\includegraphics[width=0.48\columnwidth]{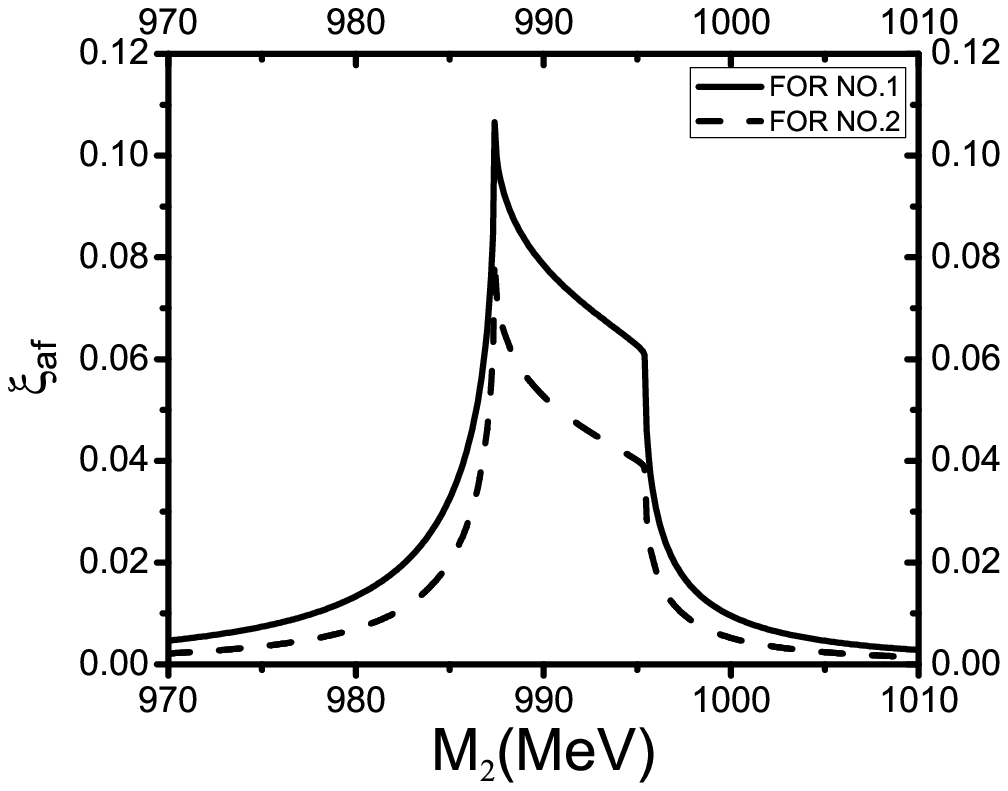}
\caption{Diagrams of \gaf intensity $\xi_{af}$ and $\xi_{fa}$ vs
two-meson invariant mass with parameter sets No.1 and No.2 of
Table~\ref{tab-2}.}\label{cho2}
\end{center}
\end{figure}

For two sets of parameters given in Table.\ref{tab-2}, we can see
that the set No.1 is close to the SND values in Table.\ref{tab-1}.
The set No.2 changes the not well measured $g_{aKK}$ and $g_{fKK}$
in their experimental uncertainties. We plot the corresponding
diagrams of \gaf intensities $\xi_{af}$ and $\xi_{fa}$ vs two-meson
invariant mass $M_2$ as shown in the Fig.\ref{cho2}. The two sets of
parameters give almost identical $\xi_{fa}$ but very different
$\xi_{af}$.

\section{Possibility of measuring $\xi_{af}$ from $\chi_{c1}\to\pi^{0}\ga \to \pi^0 \gf\to\pi^0\pi\pi$}
\label{s5}

\begin{figure}[htbp] \vspace{-0cm}
\begin{center}
\includegraphics[width=0.55\columnwidth]{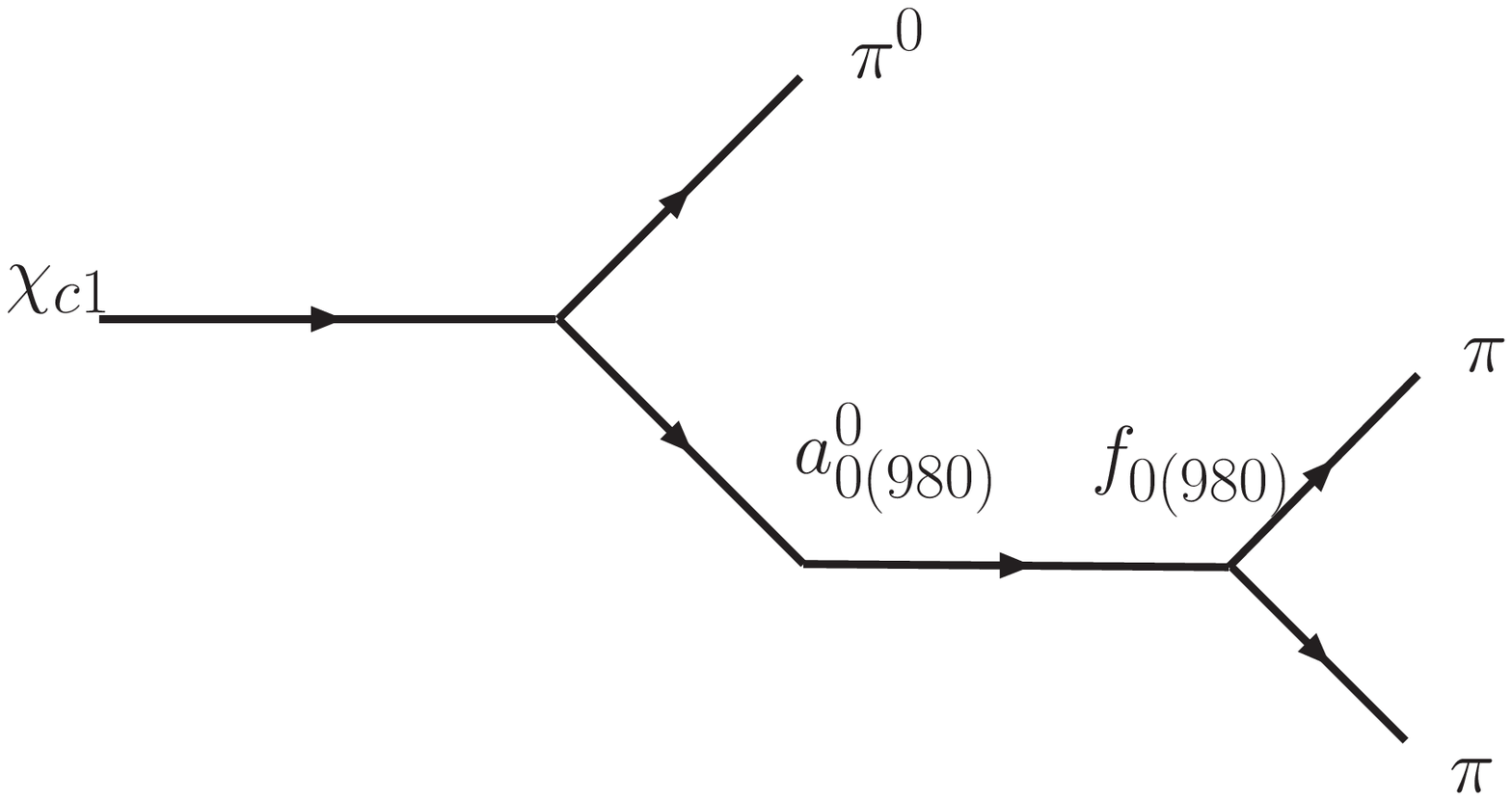}
\caption{Feynman diagram for $\chi_{c1}\to\pi^{0}\ga \to \pi^0
\gf\to\pi^0\pi\pi$}\label{f3}
\end{center}
\end{figure}

The Feynman diagram for the reaction $\chi_{c1}\to\pi^{0}\ga \to
\pi^0 \gf \to \pi^0\pi\pi$ is shown in Fig \ref{f3}.

The invariant amplitude for this reaction is
\begin{equation}
{\cal{M}}_{\chi_{c1}\to\pi^{+}\pi^{-}\pi^{0}}=g_{\chi_{c1}\ga\pi^{0}}\varepsilon^{\mu}_{\chi_{c1}}(p_{\pi^0}-p_{\gf})
\frac{D_{af}}{D_{f}D_{a}}\sqrt{3}g_{\gf\pi^0\pi^0}
\end{equation}

The coupling constant $g_{\chi_{c1}\ga\pi^{0}}$ can be determined by
the reaction $\chi_{c1}\to a^{\pm}_{0(980)}\pi^{\mp}$.
\begin{equation}
|g_{\chi_{c1}\ga\pi^{0}}|^{2}=|g_{\chi_{c1}
a^{+}_{0(980)}\pi^{-}}|^{2}=\frac{12\pi
M_{\chi_{c1}}^{2}\Gamma_{\chi_{c1}}Br_{(\chi_{c1}\to a^+_{0(980)}
\pi^-)}}{|\vec{ p}_{a^+_{0(980)}}|^{3}}.
\end{equation}

\begin{figure}[htbp] \vspace{-0cm}
\begin{center}
\includegraphics[width=0.6\columnwidth]{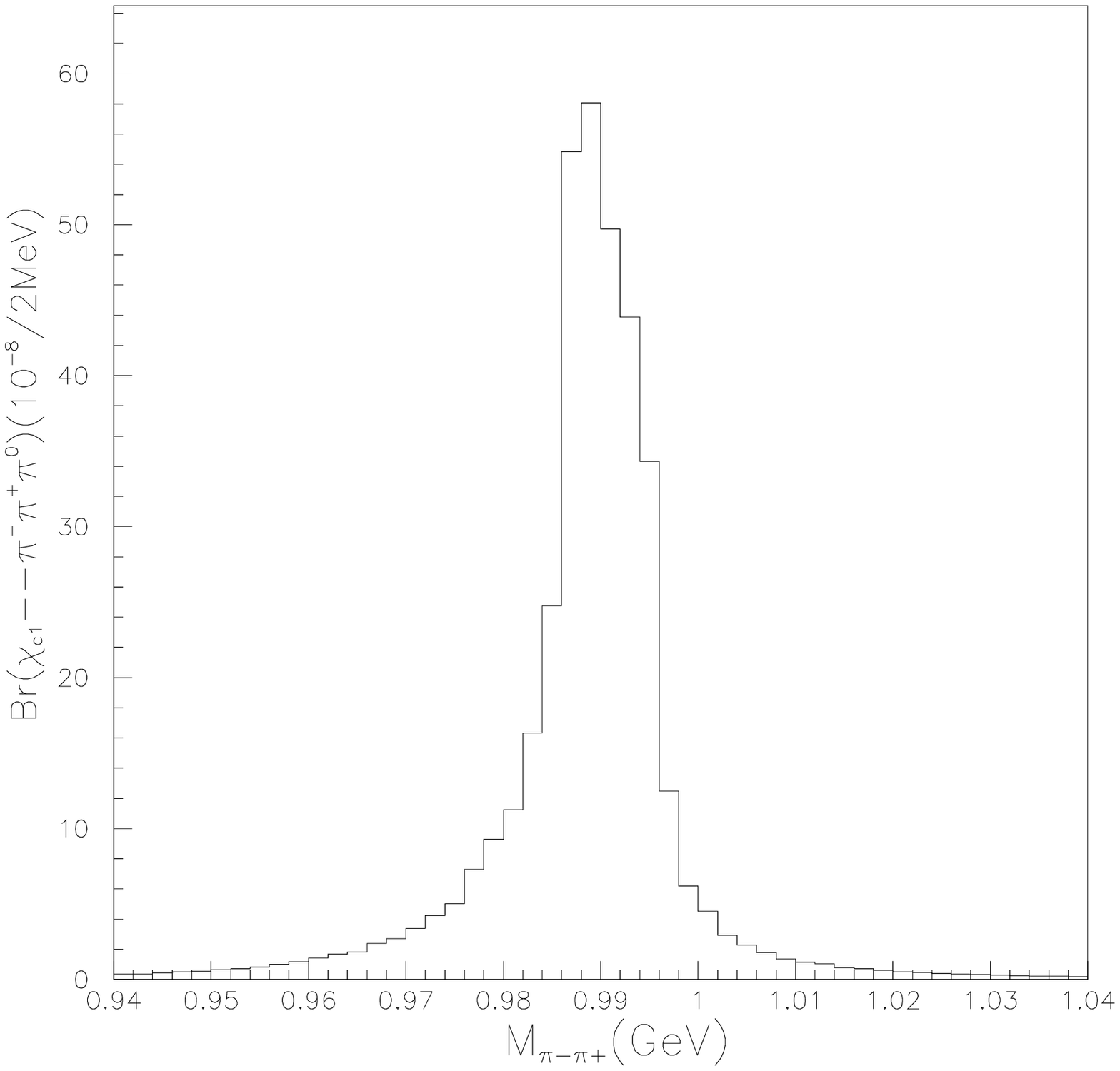}
\caption{$\pi^{+}\pi^{-}$ invariant mass spectrum for
$\chi_{c1}\to\pi^{0}\ga \to \pi^0 \gf \to \pi^0 \pi^+\pi^-$
}\label{fg}
\end{center}
\end{figure}

According to Ref.~\cite{pdg06}, $M_{\chi_{c1}}=3510.66$ MeV,
$\Gamma_{\chi_{c1}}=0.89$ MeV, and
$Br_{\chi_{c1}\to\eta\pi^+\pi^-}=(5.2\pm 0.6)\times 10^{-3}$. The
$\chi_{c1}\to a^{\pm}_{0(980)}\pi^{\mp}$ gives the dominant
contribution of about 75.1\%~\cite{cleo}. So we have
$Br_{(\chi_{c1}\to a^+_{0(980)} \pi^-)}\simeq 0.0052\times
0.751/2\simeq 2\times 10^{-3}$. From the formulae of
Eqs.(\ref{1}-\ref{6}) and parameters of setting No.H listed in Table
1 , we can calculate the $Br_{\chi_{c1}\to\pi^{0}\ga \to \pi^0 \gf
\to \pi^0 \pi\pi}$. It is $4.6\times 10^{-6}$ and the invariant mass
spectrum of $\pi^+\pi^-$ for $\chi_{c1}\to\pi^{0}\ga \to \pi^0
\gf\to\pi^0\pi^+\pi^-$ is shown in Fig.\ref{fg}.

The branching ratio of reaction $\psi_{2s}\to\gamma\chi_{c1}$ is
$0.088$~\cite{pdg06}. At the upgraded Beijing electron positron
collider with BESIII detector, about $3.2\times 10^{9}$ $\psi_{2s}$
events and hence about $2.8\times 10^{8}$ $\chi_{c1}$ events can be
collected per year. From the branching ratio of $\chi_{c1}\to\pi^0
\pi\pi=4.6\times10^{-6}$, more than $1200$ events are expected to be
collected. Considering the reconstruction efficiency of 30\%, more
than 300 events can be reconstructed for this channel. Since all
these events should concentrate in a narrow region of about 8 MeV
around 990 MeV in the $\pi^+\pi^-$ invariant mass spectrum, the
narrow peak should be easily observed, hence the $\xi_{af}$ should
be able to be measured by this reaction.

\section{Summary}
\label{s6}

Various processes have been proposed previously to study \gaf
through $\gf\to\ga$ transition. In this article we investigate in
detail the difference between $\ga\to\gf$ and $\gf\to\ga$
transitions. Two corresponding mixing intensities $\xi_{af}$ and
$\xi_{fa}$ are defined. It is found that besides the common
numerator, the $\xi_{af}$ has additional dependence on the
parameters of $\gf$ while the $\xi_{fa}$ has additional dependence
on the parameters of $\ga$. Therefore to measure $\ga\to\gf$
transition in addition to the $\gf\to\ga$ transition will be very
useful for pinning down these parameters. We examine the possibility
of measuring the \gaf from $\chi_{c1}\to\pi^0 \pi\pi$ for $\xi_{af}$
at the upgraded Beijing Electron Positron Collider with BESIII
detector and find it is feasible.

\bigskip
\noindent {\bf Acknowledgements}  This work is partly supported by
the National Natural Science Foundation of China (NSFC) and by the
Chinese Academy of Sciences under project No. KJCX3-SYW-N2.

\end{document}